\documentclass[12pt]{article}
\usepackage{amsmath,amssymb}
\usepackage{graphicx}
\newcommand{\eqdef}{\stackrel{\text{def}}{=}}

\newcommand{\n}{\nonumber\\}
\newcommand{\bm}{\boldsymbol}
\newcommand{\ignore}[1]{}
\numberwithin{equation}{section}
\newcommand{\Romannumeral}[1]{\uppercase\expandafter{\romannumeral#1}}

\newcommand{\II}{\text{\Romannumeral{2}}}

\newcommand{\tn}[1]{\tiny #1}
\newtheorem{theo}{\bf Theorem}[section]

\newtheorem{rema}[theo]{\bf Remark}

\newtheorem{cond}[theo]{\bf Condition}
\newcommand{\cH}{\mathcal{H}}
\newcommand{\cP}{\mathcal{P}}
\newcommand{\cT}{\mathcal{T}}
\newcommand{\cE}{\mathcal{E}}

\allowdisplaybreaks[4]

\begin{document}

\baselineskip=20pt

%%%%%%%%%%%%%%%%%%%%%%%%%%%%%%%%%%%%%%%%%%%%%%%%%%%%%%%%%%%%
%                                                          %
%  Title page                                              %
%                                                          %
%%%%%%%%%%%%%%%%%%%%%%%%%%%%%%%%%%%%%%%%%%%%%%%%%%%%%%%%%%%%
\newcommand{\preprint}{
\vspace*{-20mm}
   \begin{flushright}\normalsize   
%     {\tt arXiv:2212.xxxxx[quant-ph]}\\
%     December 2022
  \end{flushright}}
\newcommand{\Title}[1]{{\baselineskip=26pt
  \begin{center} \Large \bf #1 \\ \ \\ \end{center}}}
\newcommand{\Author}{\begin{center}
  \large \bf  Ryu Sasaki\end{center}}
\newcommand{\Address}{\begin{center}
      Department of Physics, Tokyo University of Science,
     Noda 278-8510, Japan
   \end{center}}
\newcommand{\Accepted}[1]{\begin{center}
  {\large \sf #1}\\ \vspace{1mm}{\small \sf Accepted for Publication}
  \end{center}}

\preprint
\thispagestyle{empty}

\Title{Quantum vs Classical Birth and Death Processes; Exactly Solvable Examples  }

\Author

\Address
\vspace{1cm}

\begin{abstract}
A coinless quantisation procedure  of continuous and discrete time Birth and Death (BD) processes
is presented. The quantum Hamiltonian $\cH$ is derived by similarity transforming 
the matrix $L$ describing the BD equation in terms of the square root of the stationary (reversible)
distribution. The quantum and classical systems share the entire eigenvalues and the eigenvectors are
related one to one. When the birth rate $B(x)$ and the death rate $D(x)$ are chosen to be the
coefficients of the difference equation governing the orthogonal polynomials of Askey scheme, 
the quantum system is exactly solvable.  The eigenvectors are the orthogonal polynomials themselves and
the eigenvalues are given analytically. Many examples are periodic since their eigenvalues are all  integers, 
or all  integers for integer parameters. The situation is very similar to the
exactly solvable one dimensional quantum mechanical systems.
These  exactly solvable Markov chains contain many adjustable free parameters 
which could be helpful for various simulation purposes.
\end{abstract}

%%%%%%%%%%%%%%%%%%%%%%%%%%%%%%%%%%%%%%%%%%%%%%%%%%%%%%%%%%%%%%%
%                                                             %
%  1. Introduction                                            %
%                                                             %
%%%%%%%%%%%%%%%%%%%%%%%%%%%%%%%%%%%%%%%%%%%%%%%%%%%%%%%%%%%%%%%
\section{Introduction}
\label{sec:intro}

Constructing the quantum counterparts of various classical stochastic processes,  in particular,
random walks on graphs,  has been an exciting hot topic in computer science, 
solid state physics,  polymer science, astronomy and mathematics. 
For physical systems in the continuous space-time the canonical quantisation is well established and 
the passage from quantum to classical regime is achieved by taking the limit of 
vanishing Planck's constant $\hbar\to0$.
 In finite discrete spaces, however, no orthodox quantisation procedure is known and  there is a plethora
of quantisation schemes reflecting the diversity of intended applications.
The prevailing pattern of quantisation is the discrete time version 
\cite{watrous, aharonov, ambainis, ambainis2, nayak, konno, kempe, szegedy, childs2, venegas} with coins, 
which are the additional degrees of freedom at the quantum level.
Sometimes the  terms like the chirality or the helicity are used. 
That means there are as many types of quantisation as the design and the number of the coins.
Among the coinless versions \cite{meyer, patel} some use the parity, {\em i.e.} the even and oddness of the 
coordinates instead of the coins.
Of course, there are many continuous time versions \cite{farhi, childs, kempe, mulken, solenov, chase, xu1, xu2, mulken2},
which naturally fit for the Schr\"odinger equations. Another scheme for  quantisation of stochastic processes is 
the open quantum systems \cite{attal} in which the density matrices play the central role.

Among these publications the classical-quantum correspondence is mostly based on the 
adjacency  matrix $A$ of the classical walk or the Markov chain. 
Since $A$ is a real symmetric matrix, the corresponding quantum Hamiltonian $H$ 
is chosen to be proportional to $A$,
or  the unitary discrete time translation operator $U$ is assembled by certain exponentiation of $A$.
Since the symmetry of the adjacency matrix is essential for the hermiticity of the quantum Hamiltonian,
the quantisation of most non-symmetric stochastic processes, {\em e.g.}
general Markov chain with the fundamental transition probability $K_{x\,y}\neq K_{y\,x}$, 
requires a different paradigm.
The Birth and Death (BD) processes \cite{feller} belong to this non-symmetric category, too.
Probably this is the reason why I have found only a few papers discussing quantum BD
processes \cite{grunbaum, ho}.

In this paper I present a quantisation scheme of the continuous  and discrete time BD processes {\em without a coin}.
The quantisation of reversible (detailed balanced) Markov chains \cite{aldous}
without a coin is described in a separate publication \cite{qcMarkov}.
Since Schr\"odinger equations or the pure state quantum mechanics are time reversal invariant,
a natural choice is the quantisation of reversible Markov chains.
The general BD processes are reversible, too.
These formulations are general and universal in the sense they apply to all processes 
with proper qualifications on finite or semi-infinite discrete base spaces.
Moreover, the classical-quantum correspondence is unique in the sense that the entire 
eigenvalues and eigenvectors are shared by the classical and quantum systems,
or they correspond one to one by simple transformation rules.
There are  plenty examples of exactly solvable classical continuous \cite{KarMcG, KarMcG2, bdsol} and discrete 
time \cite{dtbd} birth and death processes and reversible Markov chains \cite{os39},  
whose eigenvalues and eigenvectors are obtained analytically.
In all these exactly solvable examples, the eigenvectors are the hypergeometric orthogonal polynomials
of a discrete variable belonging to the Askey scheme \cite{askey, ismail, kls}.
For example, the Krawtchouk, Hahn, Racah polynomials etc and their $q$-versions for the finite base spaces 
and the Charlier, Meixner and little $q$-Jacobi polynomials,  etc. for the semi-infinite base spaces.
They contain many free parameters.
For the continuous time BD processes, 
the quantum Hamiltonian is a positive semi-definite real symmetric
tri-diagonal matrix of finite or infinite dimensions composed of the birth and death rates.
The unitary matrices governing the finite time translation of the quantum discrete time 
BD  processes and  Markov chains are also constructed explicitly by the birth and death rates
or the transition probability matrices and the reversible distributions of the classical Markov chains.

This paper is organised as follows. In section two the setting of general classical continuous time 
BD processes with the
birth rate $B(x)$ and death rate $D(x)$ is recapitulated \cite{bdsol}.
The matrix $L$ \eqref{LBDdef}--\eqref{bdproconv} 
describing the BD equation \eqref{BDeq} is similarity transformed to 
a {\em real symmetric positive semi-definite} matrix $\mathcal{H}$ \eqref{LBDHrel}--\eqref{Hdef3} in terms of
$\phi_0(x)$ \eqref{phi0def}, which is the square root of 
the  corresponding reversible (stationary) distribution $\pi(x)$ \eqref{pidef}.
By another similarity transformation of $\mathcal{H}$ in terms of $\phi_0(x)$, a third tri-diagonal matrix
$\widetilde{\mathcal H}$ \eqref{htdef}--\eqref{htdef3} is introduced, whose eigenvalue problem is a
second order difference equation \eqref{hteig2} of the eigenvectors.
{\bf Theorem \ref{theo1}} states that the general classical BD problem is reduced to the solution of the
eigenvalue problem of the quantum Hamiltonian $\mathcal H$ \eqref{Hdef1}--\eqref{Hdef3}.
In section three the setting of general discrete time BD processes is introduced.
The evolution of the general discrete time  BD process is described in {\bf Theorem \ref{theo2}}.
Now it is clear that the evolution of continuous and discrete time BD processes are very closely related 
to each other.  The evolution and the branching probability amplitudes of  quantum BD processes of
continuous and discrete time BD processes are stated in {\bf Theorem \ref{theo3}} in section four. 
The classical and quantum BD processes share the entire eigenvalues and the eigenvectors are
related by the multiplication of $\phi_0(x)$ \eqref{phindef}--\eqref{phindef2}.
The main issue of the exactly solvable quantum BD processes is presented in section five.
For 16 choices of the birth rate $B(x)$ and death rate $D(x)$, which are the coefficients of 
the difference equation \eqref{difeq} governing the hypergeometric orthogonal polynomials of 
Askey scheme \cite{askey, kls}, the eigenvalues are explicitly known and the eigenvectors 
are the polynomials themselves.
The identifications of the eigenvectors and the orthogonal polynomials are stated in \S\ref{sec:cBDex}
and \S\ref{sec:dBDex} for the continuous and discrete time BD. 
The formulas of the probability amplitudes etc in terms of the orthogonal polynomials are shown in \S\ref{sec:exform}.
The explicit data of $B(x)$, $D(x)$, $\cE(n)$ etc of five exactly solvable BD cases are demonstrated in \S\ref{sec:data}.
The data for the remaining 11 cases are 
listed in \cite{bdsol}.
It is pointed out that in some cases all the eigenvalues are integers and the quantum systems are periodic.
 Section six  is for many comments. 
Basic definitions related to the ($q$)-hypergeometric functions are supplied in Appendix for self-consistency.

%%%%%%%%%%%%%%%%%%%%%%%%%%%%%%%%%%%%%%%%%%%%%%%%%%%%%%%%%%%%%%%
%                                                             % 
%  2. Continuous time birth and death processes                                         %
%                                                             %
%%%%%%%%%%%%%%%%%%%%%%%%%%%%%%%%%%%%%%%%%%%%%%%%%%%%%%%%%%%%%%%
\section{Continuous time birth and death processes}
\label{sec:cBD}
Birth and Death (BD) processes are a special branch of the general stationary Markov processes (chains) on graphs,
having the nearest neighbour interactions only.
Let us denote the set of vertices of the graph $G$ by $V$, either finite or semi-infinite:
\begin{equation}
V=\{0,1,\ldots,N\}: \quad \text{finite},\qquad V
=\mathbb{Z}_{\ge0}: \quad \text {semi-infinite}.
\label{GVx}
\end{equation}
For analytic treatments  I  use $x,y,..$ as representing the the vertices of V,  $x,y\in{V}$ 
rather than the conventional $n$ and $m$ etc.
Let us denote the birth rate at population $x$ by $B(x)>0$ going from $x$ to $x+1$ and the death rate by $D(x)>0$
going from $x$ to $x-1$.
In order to keep non-negative populations and the maximal population $N$ in the semi-infinite case, the following boundary conditions are imposed,
\begin{equation}
D(0)=0,\quad B(N)=0:\  (\text{only  for a finite case}),
\label{bdcond1}
\end{equation}
which are  called the reflecting boundary conditions.
Let $\mathcal{P}(x;t)\ge0$ be the probability distribution  over $V$ at time $t$.
The  continuous time evolution of the probability distribution is governed by the following differential equation:
\begin{align}
\frac{\partial}{\partial t}\mathcal{P}(x;t)%
&=-(B(x)+D(x))\mathcal{P}(x;t)+B(x-1)\mathcal{P}(x-1;t)+D(x+1)\mathcal{P}(x+1;t),
\label{BDeq}\\
&=(L\mathcal{P})(x;t)=\sum_{y\in{V}}L_{x\,y}\mathcal{P}(y;t),
\quad  \sum_{x\in{V}} \mathcal{P}(x;t)=1,
\label{bdeqformal}
\end{align}
Here  $L$ is  a  {\em tri-diagonal} matrix on $\ell^2(V)$,
\begin{align}
&{L}_{x+1\,x}=B(x),\ \  {L}_{x-1\,x}=D(x),\ {L}_{x\,x}=-B(x)-D(x), \ \
L_{x\,y}=0,\quad |x-y|\ge2,
\label{LBDdef}
%\\
%%
%&L=\n
%%
%&\left(
%\begin{array}{cccccc}
%-B(0)  & D(1)  &   0& \cdots&\cdots&0\\
%B(0)  &-B(1)-D(1)  &D(2) &0&\cdots&\vdots  \\
%\!\!0  &  B(1)  &  -B(2)-D(2)&D(3)&\cdots&\vdots\\
%\!\!\vdots&\cdots&\cdots&\cdots&\cdots&\vdots\\
%\!\!\vdots&\cdots&\cdots&\cdots&\cdots&0\\
%\!\!0&\cdots&\cdots&B(N\!\!-\!\!2)&-B(N\!\!-\!\!1)\!-\!D(N\!\!-\!\!1)&D(N)\\
%\!\!0&\cdots&\cdots&0&B(N\!-\!1)&-D(N)
%\end{array}
%\right),\nonumber
\end{align}
satisfying the condition
\begin{equation}
\sum_{x\in{V}}{L}_{x\,y}=0.
\label{bdproconv}
\end{equation}
This ensures the conservation of probability, that is, the condition
$\sum_{x\in{V}} \mathcal{P}(x;t)=1$  is preserved by the time evolution \eqref{bdeqformal}.
It should be stressed that  in the continuous time formulation, 
the {\em time scale is completely arbitrary}. In other words, the overall scale of 
the birth/death ratio $B(x)$ and $D(x)$ can be multiplied by any factor depending on various parameters.

Let us introduce a positive function 
$\phi_0(x)$ on ${V}$ and a {\em diagonal matrix} $\Phi$ consisting of $\phi_0(x)$ 
by the ratios of $B(x)$ and $D(x+1)$,
\begin{align} 
  &\phi_0(0)\eqdef1,\quad \phi_0(x)\eqdef\sqrt{\prod_{y=0}^{x-1}\frac{B(y)}{D(y+1)}}
  \Leftrightarrow  \frac{\phi_0(x+1)}{\phi_0(x)}=\frac{\sqrt{B(x)}}{\sqrt{D(x+1)}},\quad x\in{V}.
  \label{phi0def}\\
& \Phi_{x\,x}=\phi_0(x),\quad \Phi_{x\,y}=0, \quad x\neq y.
\label{Phidef}
\end{align}
For semi-infinite cases, $B(x)$ and $D(x)$ must be restricted so that $\phi_0(x)$ is square summable,
\begin{equation}
\sum_{x\in{V}}\phi_0(x)^2<\infty,\qquad \phi_0\in\ell^2(V).
\end{equation}
By a  similarity transformation 
in terms of $\Phi$, let us introduce another {\em tri-diagonal} matrix  ${\mathcal H}$ on $\ell^2(V)$,
\begin{align} 
   &\hspace{50mm}\mathcal{H}\eqdef-\Phi^{-1}{L}\Phi,
   \label{LBDHrel} \\
& \mathcal{H}_{x\,x}=-{L}_{x\,x}=B(x)+D(x),\qquad \mathcal{H}_{x\,y}=0, \quad |x-y|\ge2,
\label{Hdef1}\\
&\mathcal{H}_{x+1\,x}=-\phi_0(x+1)^{-1}{L}_{x+1\,x}\phi_0(x)
=-\phi_0(x+1)^{-1}B(x)\,\phi_0(x)
=-\sqrt{B(x)D(x+1)},
\label{Hdef2}\\
&\mathcal{H}_{x-1\,x}=-\phi_0(x-1)^{-1}{L}_{x-1\,x}\phi_0(x)
=-\phi_0(x-1)^{-1}\,D(x)\phi_0(x)
=-\sqrt{B(x-1)D(x)}.
\label{Hdef3}
\end{align}
The matrix $\mathcal{H}$ is the {\em quantum Hamiltonian} corresponding to the classical continuous time
birth and death process \eqref{BDeq} as it is 
{\em real symmetric}  and  {\em positive semi-definite} as can be seen clearly 
by the following factorisation in terms of an upper triangular matrix $\mathcal{A}$,
\begin{align}
\mathcal{H}&={}^t\!\!\mathcal{A}\mathcal{A}\ \Rightarrow \mathcal{H}
={}^t\mathcal{H},
\label{Hsym}\\
 \mathcal{A}_{x\,x}&\eqdef\sqrt{B(x)},\quad 
\mathcal{A}_{x\,x+1}\eqdef-\sqrt{D(x+1)},\quad \mathcal{A}_{x\,y}\eqdef0,\ \text{otherwise},
\quad x, y\in{V},
\label{Adef}
\end{align}
in which ${}^t\!\!\mathcal{A}$ is the transposed matrix of $\mathcal{A}$.
The zero mode (eigenvector) of $\mathcal{A}$ and $\mathcal{H}$
 is  $\phi_0(x)$
\begin{equation}
0=(\mathcal{A}\phi_0)(x)=\sqrt{B(x)}\phi_0(x)-\sqrt{D(x+1)}\phi_0(x+1)
\ \Rightarrow (\mathcal{H}\phi_0)(x)=0.
\label{Azero}
\end{equation}
The factorisation guarantees the reality and non-negativeness of the eigenvalues of $\mathcal{H}$,
\begin{equation}
\mathcal{E}(0)=0,\quad
\mathcal{E}(n)>0,\qquad n\in{V}\backslash\{0\},
\label{posE}
\end{equation}
and the orthogonality of the eigenvectors  of $\mathcal{H}$, since the  simpleness of
the spectrum is due to its tri-diagonality. 
The positive semi-definiteness of $\cH$ \eqref{posE} means {\em negative semi-definiteness}
of $L$ \eqref{bdeqformal}, which is essential for the stability of the system.

Next let us introduce another tri-diagonal matrix $\widetilde{\mathcal H}$ by the similarity transformation
in terms of $\Phi$,
\begin{align} 
     & \hspace{3cm}  \widetilde{\mathcal H}\eqdef\Phi^{-1} \mathcal{H} \Phi \ \Leftrightarrow
   \widetilde{\mathcal H}_{x\,y}=\phi_0(x)^{-1}{\mathcal H}_{x\,y}\phi_0(y),
   \label{htdef}\\
 &\widetilde{\mathcal H}_{x\,x+1}=-B(x),\   \widetilde{\mathcal H}_{x\,x-1}=-D(x),\ 
\widetilde{\mathcal H}_{x\,x}=B(x)+D(x), \ 
\widetilde{\mathcal H}_{x\,y}=0, |x-y|\ge2.
\label{htdef2}\\
&\hspace{3cm} \Longrightarrow \sum_{y\in{V}}\widetilde{\mathcal H}_{x\,y}=0.
\label{htdef3}
\end{align}
The eigenvalue problem of $\widetilde{\mathcal H}$
\begin{align} 
(\widetilde{\mathcal H}v_n)(x)=\sum_{y\in{V}}\widetilde{\mathcal H}_{x\,y}v_n(y)=\mathcal{E}(n)v_n(x),
\quad x,n\in{V}.
\label{hteig1}
\end{align}
takes a simple form of a second order difference equation for $v_n(x)$,
\begin{align}
&B(x)\bigl(v_n(x)-v_n(x+1)\bigr)+D(x)\bigl(v_n(x)-v_n(x-1)\bigr)=\mathcal{E}(n)v_n(x),
\quad x,n\in{V}.
\label{hteig2}
\end{align}
It should be stressed that for finite BD processes, $\{v_n(x)\}$ are eigenvectors. 
But for infinite systems, they might not be eigenvectors as such. For examplre
\begin{align}
\hspace{1cm} \cE(0)=0,\quad  v_0(x)\eqdef1.
\label{v0def}
\end{align}
is a solution but it is definitely not square summable $\sum_{x\in{V}}v_0(x)^2=\infty$.
Since the first component of the eigenvector of a tri-diagopnal matrix can never vanish, 
it is natural to adopt the following universal normalisation of the solutions $\{v_n(x)\}$,
\begin{equation}
v_n(0)=1,\quad n\in{V}.
\label{vnorm}
\end{equation}
This assertion is valid irrespective of the square summability of $\{v_n(x)\}$.
The eigenvectors of $\mathcal{H}$ and $L$ are
\begin{align} 
  (\mathcal{H}\phi_n)(x)&=\mathcal{E}(n)\phi_n(x),\qquad \phi_n(x)\eqdef\phi_0(x)v_n(x),
  \quad n\in{V},
  \label{phindef}\\
(L\phi_0\phi_n)(x)&=-\mathcal{E}(n)\phi_0(x)\phi_n(x), \hspace{3.7cm} n\in{V},
  \label{phindef2}
\end{align}
so long as $\{\phi_n\}$ are square summable, $\phi_n\in\ell^2(V)$,
\begin{align} 
(\phi_n,\phi_m)\eqdef \sum_{x\in{V}}\phi_n(x)\phi_m(x)
= \sum_{x\in{V}}\phi_0(x)^2v_n(x)v_m(x)
=\frac1{d_n^2}\delta_{n\,m},
 \quad n, m\in{V}.
  \label{orth}
\end{align}
Here $\phi_0(x)^2$ provides the orthogonality measure of the vectors  $\{v_n(x)\}$ 
of $\widetilde{\mathcal H}$.
The normalisation constant is positive $d_n>0$.

Let us define orthonormal vectors $\{\hat{\phi}_n(x)\}$  in $\ell^2(V)$,
\begin{equation}
\hat{\phi}_n(x)\eqdef d_n\phi_n(x)=d_n\phi_0(x)v_n(x),\quad (\hat{\phi}_n,\hat{\phi}_m)=\delta_{n\,m},
\quad n,m \in{V},
\label{normphin}
\end{equation}
and the square of the normalised zero mode
\begin{align}
\pi(x)&\eqdef \hat{\phi}_0(x)^2=d_0^2\phi_0(x)^2=d_0^2\,\prod_{y=0}^{x-1}\frac{B(y)}{D(y+1)},
\label{pidef}\\
& \sum_{x\in{V}}\pi(x)=1 \Leftarrow \frac1{d_0^2}\eqdef\sum_{x\in{V}}\prod_{y=0}^{x-1}\frac{B(y)}{D(y+1)},\end{align}
which will turn out to be the {\em stationary distribution} of the classical birth and death process.

The solution of the general classical continuous time BD  problem \eqref{BDeq} with the birth rate $B(x)$ 
and death rate $D(x)$ is reduced to the eigenvalue problem of the corresponding quantum 
Hamiltonian $\mathcal{H}$ \eqref{LBDHrel}--\eqref{Hdef3}.  It should be stressed that $\mathcal{H}$
has a very special form corresponding to the nearest neighbour interactions 
with non-vanishing elements on the diagonal, super and sub-diagonals only.
It is summarised by the following\\
\begin{theo}
\label{theo1}
By using the complete set of eigensystem $\{\mathcal{E}(n),\hat{\phi}_n(x)\}$, $x,n\in{V}$, 
of the quantum Hamiltonian ${\mathcal H}$ 
\eqref{LBDHrel}--\eqref{Hdef3}, \eqref{phindef},
the solution of the initial value problem of the classical continuous time BD  \eqref{LBDdef} 
is given by
\begin{equation}
\mathcal{P}(x;t)=\hat{\phi}_0(x)\sum_{n\in{V}}c_ne^{-\mathcal{E}(n)t}\hat{\phi}_n(x),
\label{ctbdsol1}
\end{equation}
in which $\{c_n\}$ are determined as the expansion coefficients of the initial distribution $\mathcal{P}(x;0)$,
\begin{equation}
\mathcal{P}(x;0)=\hat{\phi}_0(x)\sum_{n\in{V}}c_n\hat{\phi}_n(x) \Rightarrow
c_0=1,\ c_n=\sum_{x\in{V}}\hat{\phi}_n(x)\hat{\phi}_0(x)^{-1}\mathcal{P}(x;0),
\quad n=1,\ldots.
\label{cndef}
\end{equation}
The transition matrix from $y$ to $x$ after time $t$ is
\begin{equation}
\mathcal{P}(x,y;t)=\hat{\phi}_0(x)\hat{\phi}_0(y)^{-1}
\sum_{n\in{V}}e^{-\mathcal{E}(n)t}\hat{\phi}_n(x)\hat{\phi}_n(y).
\label{cpr1}
\end{equation}
It is straightforward to  verify the Chapman-Kolmogorov equation
\begin{equation}
\mathcal{P}(x,y;t+s)=\sum_{z\in{V}}\mathcal{P}(x,z;t)\mathcal{P}(z,y;s),
\label{chap}
\end{equation}
 by the orthogonality relation \eqref{normphin} and the completeness relation
 \begin{equation}
\sum_{n\in{V}}\hat{\phi}_n(x)\hat{\phi}_n(y)=\delta_{x\,y},\quad x,y\in{V}.
\label{phincomp}
\end{equation}
The approach to the stationary (or terminal) distribution $\pi(x)$ \eqref{pidef}  is guaranteed by the positivity of $\mathcal{E}(n)>0$
for the non-zero modes, irrespective of the initial distribution,
\begin{equation}
\lim_{t\to\infty}\mathcal{P}(x;t)=\pi(x),\quad \lim_{t\to\infty}\mathcal{P}(x,y;t)=\pi(x).
\label{piapp}
\end{equation}
Various stochastic quantities can be derived from these expressions.
\end{theo}
This was reported as {\bf Theorem 3.2} in \cite{dtbd}.

%%%%%%%%%%%%%%%%%%%%%%%%%%%%%%%%%%%%%%%%%%%%%%%%%%%%%%%%%%%%%%%
%                                                             % 
%  3. Discrete time birth and death processes                                         %
%                                                             %
%%%%%%%%%%%%%%%%%%%%%%%%%%%%%%%%%%%%%%%%%%%%%%%%%%%%%%%%%%%%%%%
\section{Discrete time birth and death processes}
\label{sec:dBD}
The  formulation of classical discrete time BD processes goes almost parallel with the continuous time case,
with the same notation of G, V, and $x,y\in{V}$ as in \eqref{GVx}.
The birth  and death probability at $x$ per unit time interval is denoted  by $0<b(x)<1$ and $0<d(x)<1$ 
satisfying the same  boundary condition as \eqref{bdcond1}
\begin{equation}
d(0)=0,\quad b(N)=0:\  (\text{only  for a finite case}),\quad 0<b(x)+d(x)<1.
\label{bdcond2}
\end{equation}
Let $\mathcal{P}^d(x;\ell)\ge0$ ($\sum_{x\in{V}}\mathcal{P}^d(x;\ell)=1$)
be the probability distribution at population $x$ over $V$ at  $\ell$-th time step.
The next step distribution  is given by
\begin{equation}
\mathcal{P}^d(x;\ell+1)=\sum_{y\in{V}}K_{x\,y}\mathcal{P}^d(y;\ell).
\label{nextstep}
\end{equation}
The initial value problem is to determine $\cP^d(x;\ell)$  for given $\cP^d(x;0)$,
\begin{equation}
\cP^d(x;\ell)=\sum_{y\in{V}}(K)^\ell_{x\,y}\cP^d(y;0),\quad \sum_{x\in{V}}\cP^d(x;0)=1,\quad \cP^d(x;0)\ge0.
\label{dinipro}
\end{equation}
Here  $K$ is  a  non-negative {\em tri-diagonal} matrix on $\ell^2(V)$,
\begin{align}
&{K}_{x+1\,x}=b(x),\ \  {K}_{x-1\,x}=d(x),\ {K}_{x\,x}=1-b(x)-d(x), \ \
K_{x\,y}=0,\quad |x-y|\ge2,
\label{KBDdef}
\end{align}
satisfying the condition
\begin{equation}
\sum_{x\in{V}}{K}_{x\,y}=1 \
\Rightarrow 
\sum_{x\in{V}}\mathcal{P}^d(x;\ell+1)=\sum_{y\in{V}}
\sum_{x\in{V}}K_{x\,y}\mathcal{P}^d(y;\ell)=\sum_{y\in{V}}\mathcal{P}^d(y;\ell)=1.
\label{Kbdproconv}
\end{equation}
In view of the quantum  version, which is time reversal invariant, the following condition  is imposed.
\begin{cond}
The system has a reversible (detail balanced) distribution, satisfying
\begin{equation}
K_{x\,y}\pi(y)=K_{y\,x}\pi(x),\qquad \sum_{x\in{V}}\pi(x)=1, \quad x,y\in{V}.
\label{revcond}
\end{equation}
\end{cond}
It is clear that $\pi(x)$ is the eigenvector of $K$ with  eigenvalue 1,
\begin{equation}
\sum_{y\in{V}}K_{x\,y}\pi(y)=\sum_{y\in{V}}K_{y\,x}\pi(x)=\pi(x).
\label{higheig}
\end{equation}
The reversibility condition \eqref{revcond} connects $\pi(x)$ with $\pi(x\pm1)$,
\begin{align}
&K_{x\,x+1}\pi(x+1)=K_{x+1\,x}\pi(x) \Rightarrow d(x+1)\pi(x+1)=b(x)\pi(x),
\label{Kpirel}\\
&\Rightarrow \pi(x)\propto \phi_0(x)^2\eqdef\prod_{y=0}^{x-1}\frac{b(y)}{d(y+1)},\quad
\frac1{d_0^2}\eqdef\sum_{x\in{V}}\prod_{y=0}^{x-1}\frac{b(y)}{d(y+1)},\quad \phi_0(0)=1,\  \phi_0(x)>0,\n
%)
&\ \quad \pi(x)= \hat{\phi}_0(x)^2\eqdef\phi_0(x)^2d_0^2>0.
\label{dpidef}
\end{align}
Here the same notation as in \eqref{pidef} is used purposefully.
Since the eigenvector $\pi(x)$ is all positive, it belongs to the maximal eigenvalue.
Perron-Frobenius theorem tells that
$K$'s spectrum is bounded by $1$ and $-1$:
\begin{equation}
-1\le\text{Eigenvalues}(K)\le1.
\label{eigbound}
\end{equation}
Since $K$ is tri-diagonal all the eigenvalues are non-degenerate.

By dividing the reversibility condition \eqref{revcond} by $\sqrt{\pi(x)\pi(y)}$, a non-negative real symmetric
tri-diagonal matrix $\cT$ is obtained,
\begin{align} 
  {\cT}_{x\,y}&\eqdef \frac1{\sqrt{\pi(x)}}K_{x\,y}\sqrt{\pi(y)}=\frac1{\sqrt{\pi(y)}}K_{y\,x}\sqrt{\pi(x)}\n
  &=\phi_0(x)^{-1}K_{x\,y}\phi_0(y)=\phi_0(y)^{-1}K_{y\,x}\phi_0(x)={\cT}_{y\,x},
  \quad {\cT}_{x\,y}=0,\ |x-y)\ge2,
  \label{Tdef}\\
{\cT}_{x+1\,x}&=\sqrt{b(x)d(x+1)},\quad {\cT}_{x-1\,x}=\sqrt{b(x-1)d(x)},\quad  {\cT}_{x\,x}=1-b(x)-d(x).
\end{align}
This is equivalent to the similarity transformation of $K$ in terms of $\Phi$,
\begin{align} 
\cT=\Phi^{-1}K\Phi,\qquad \Phi_{x\,x}=\phi_0(x),\quad \Phi_{x\,y}=0, \quad x\neq y.
\label{TKtrans}
\end{align}
The Hamiltonian   of the discrete time BD process $\cH^d$ is defined by
\begin{align} 
  &\hspace{1cm}\cH^d\eqdef \bm{1}-\cT =\Phi^{-1}\bigl(\bm{1}-K\bigr)\Phi\ \Leftrightarrow \cT=\bm{1}-\cH^d, \hspace{1cm} 
\  \quad {\cH}_{x\,y}^d=0,\ |x-y|\ge2,
  \label{Hddef}\\[2pt]
& {\cH}_{x+1\,x}^d=-\sqrt{b(x)d(x+1)},\quad {\cH}_{x-1\,x}^d=-\sqrt{b(x-1)d(x)},
\quad  {\cH}_{x\,x}=b(x)+d(x),
 \label{Hddef2}\\[2pt]
& \hspace{3cm} 0\le\text{Eigenvalues}(\cH^d)\le2,
\label{eigbound2}
\end{align}
in which $\bm{1}$ is the identity operator in $\ell^2(V)$. 
That is, $\cH^d$ is positive semi-definite, too.
It is well known that for a discrete time Markov chain $K$, $K-\bm{1}$ corresponds 
to the Laplacian \cite{woess}.

Since $\cH^d$ \eqref{Hddef} has the same form as $\cH$, the continuous time BD Hamiltonian 
$\cH$  \eqref{Hdef1}--\eqref{Hdef3},  the solutions of the eigenvalue problems have the same form;
\begin{align}
 (\cH^d\phi_n)(x)&=\cE^d(n)\phi_n(x),\\
 (\cT\phi_n)(x)&=(1-\cE^d(n))\phi_n(x),\\
 (K\phi_0\phi_n)(x)&=\kappa(n)\phi_0(x)\phi_n(x),\qquad \kappa(n)\eqdef1-\cE^d(n).
\end{align}
The spectral representation of $K$ is
\begin{equation}
K_{x\,y}=\hat{\phi}_0(x)\sum_{n\in{V}}\kappa(n)\hat{\phi}_n(x)\hat{\phi}_n(y)\hat{\phi}_0(y)^{-1}.
\end{equation}
Corresponding to {\bf Theorem \ref{theo1}} the solutions of the discrete time BD process 
are summarised by the following 
\begin{theo}
\label{theo2}
By using the complete set of eigensystem $\{\mathcal{E}^d(n),\hat{\phi}_n(x)\}$, $x,n\in{V}$, 
of the quantum Hamiltonian ${\mathcal H}^d$ 
\eqref{Hddef}--\eqref{Hddef2},
the solution of the initial value problem of the classical discrete time BD  \eqref{LBDdef} 
is given by
\begin{equation}
\mathcal{P}^d(x;\ell)=\hat{\phi}_0(x)\sum_{n\in{V}}c_n\bigl(\kappa(n)\bigr)^\ell\hat{\phi}_n(x),
\label{ctbdsol2}
\end{equation}
in which $\{c_n\}$ are  given in \eqref{cndef}.
The transition probability matrix from $y$ to $x$ after $\ell$  steps is
\begin{equation}
\mathcal{P}^d(x,y;\ell)=\hat{\phi}_0(x)\hat{\phi}_0(y)^{-1}
\sum_{n\in\mathcal{X}}(\kappa(n))^\ell\hat{\phi}_n(x)\hat{\phi}_n(y).
\label{ltrpr}
\end{equation}
In the absence of the eigenvalue $-1$, the stationary distribution $\pi(x)$ 
\eqref{revcond}, \eqref{dpidef} is attained  after large steps,
\begin{equation}
\lim_{\ell\to\infty}\mathcal{P}^d(x;\ell)=\pi(x),\qquad \lim_{\ell\to\infty}\mathcal{P}^d(x,y;\ell)=\pi(x).
\end{equation}

\end{theo}
%%%%%%%%%%%%%%%%%%%%%%%%%%%%%%%%%%%%%%%%%%%%%%%%%%%%%%%%%%%%%%%
%                                                             %
%  4. Quantum  birth and death processes                                           %
%                                                             %
%%%%%%%%%%%%%%%%%%%%%%%%%%%%%%%%%%%%%%%%%%%%%%%%%%%%%%%%%%%%%%%
\section{Quantum  birth and death processes}
\label{sec:qBD}
The quantum  BD process, continuous and discrete time, is described in the Hilbert space $\ell^2(V)$ 
with the orthonormal basis
\begin{equation*}
|x\rangle,\quad \langle y|x\rangle=\delta_{y\,x},\quad x,y\in{V}.
\end{equation*}
In the conventional vector notation it is
\begin{equation}
|x\rangle\equiv e_x=\vspace{-4pt}
\begin{array}{ccccccc}
 & \!\! \tn{0} & \tn{\cdots}& \tn{x}&\tn{\cdot}&\tn{\cdots}& \\
{}^t(  & \!\! 0 &  \cdots& 1& 0&\cdots&)\\
& & & & & &
\end{array}
\ \Rightarrow \sum_{x\in{V}}|x\rangle\langle x|=\bm{1},
\label{xcomp}
\vspace{-10pt}
\end{equation}
in which $\bm{1}$ is the identity matrix in $\ell^2(V)$.
Another orthonormal basis is the normalised eigenvectors of the Hamiltonian $\mathcal{H}$, 
$\{\hat{\phi}_n\}$, $(\hat{\phi}_m,\hat{\phi}_n)=\delta_{m\,n}$, $n,m\in{V}$ \eqref{phindef}. 
By using the notation $|x\rangle$, it can be expressed by $\|n\rangle\!\rangle$,
\begin{align}
 &\|n\rangle\!\rangle\eqdef\sum_{x\in{V}}\hat{\phi}_n(x)|x\rangle \ 
 \Longleftrightarrow \
\langle x\|n\rangle\!\rangle=\hat{\phi}_n(x),
\qquad \langle\!\langle m\|n\rangle\!\rangle=\delta_{m\,n},\quad n,m\in{V},\\
&\hspace{8cm}  \Longrightarrow \sum_{n\in{V}}\|n\rangle\!\rangle\langle\!\langle n\|=\bm{1}.
\end{align}
This is the completeness relation \eqref{phincomp}
\begin{equation}
\sum_{n\in{V}}\langle x\|n\rangle\!\rangle\langle\!\langle n\|y\rangle
=\sum_{n\in{V}}\hat{\phi}_n(x)\hat{\phi}_n(y)=\delta_{x\,y},\quad x,y\in{V}.
\label{ncomp}
\end{equation}
In a sense they are `dual' to each other, $\{|x\rangle\}$ specify the population (coordinates, vertices) and 
$\{\|n\rangle\!\rangle\}$ specify the eigenlevels (eigenvalues).
The spectral representations of the Hamiltonians $\mathcal{H}$  and $\cH^d$ read
\begin{align}
\mathcal{H}&=\sum_{n\in{V}}\mathcal{E}(n)\|n\rangle\!\rangle\langle\!\langle n\|\ \
\Longleftrightarrow \ \mathcal{H}_{x\,y}=\langle x|\mathcal{H}|y\rangle\ =\sum_{n\in{V}}\mathcal{E}(n)\hat{\phi}_n(x)\hat{\phi}_n(y),
\label{Hspec}\\
\mathcal{H}^d&=\sum_{n\in{V}}\cE^d(n)\|n\rangle\!\rangle\langle\!\langle n\|\
\Longleftrightarrow \ \mathcal{H}^d_{x\,y}=\langle x|\mathcal{H}^d|y\rangle
=\sum_{n\in{V}}\mathcal{E}^d(n)\hat{\phi}_n(x)\hat{\phi}_n(y).
\label{Hdspec}\
\end{align}
In contrast to the continuous time BD equation \eqref{BDeq}
a quantum state $|\psi(t)\rangle$ evolves according to the Schr\"odinger equation
\begin{equation}
i\frac{\partial}{\partial t}|\psi(t)\rangle=\mathcal{H}|\psi(t)\rangle.
\label{Scheq}
\end{equation}
The evolution of the discrete time system is governed by a unitary matrix $U\in\ell^2(V)$ 
which is the exponentiation of $\cH^d$,
\begin{equation}
|\psi^d(\ell+1)\rangle=U|\psi^d(\ell)\rangle,\quad U\eqdef e^{-i\cH^d}
=\sum_{n\in{V}}e^{-i\cE^d(n)}\|n\rangle\!\rangle\langle\!\langle n\|\ .
\label{Udef}
\end{equation}

As seen above the classical and quantum  BD processes are the head and tail of the same coin.
The normalised general initial states $|\psi(0)\rangle$,  $|\psi^d(0)\rangle$ can be represented as
\begin{align}
|\psi(0)\rangle&=\sum_{z\in{V}}\varphi(z)|z\rangle,\quad \ \varphi(z)\in\mathbb{C},
\quad  \ \sum_{z\in{V}}|\varphi(z)|^2=1,
\label{inic}\\
|\psi^d(0)\rangle&=\sum_{z\in{V}}\varphi^d(z)|z\rangle,\quad \varphi^d(z)\in\mathbb{C},
\quad \sum_{z\in{V}}|\varphi^d(z)|^2=1.
\label{inid}
\end{align}

The following Theorem corresponds to {\bf Theorem \ref{theo1}, \ref{theo2}}.
\begin{theo}
\label{theo3}
In terms of  the complete sets of eigensystem $\{\mathcal{E}(n),\hat{\phi}_n(x)\}$, 
$\{\mathcal{E}^d(n),\hat{\phi}_n(x)\}$, $x,n\in{V}$, 
of the quantum Hamiltonians ${\mathcal H}$, $\cH^d$, 
the solution of the initial value problem of the Schr\"odinger equation \eqref{Scheq} is
\begin{align} 
 |\psi(t)\rangle=e^{-i\mathcal{H}t} |\psi(0)\rangle
&=\sum_{n\in{V}}e^{-i\mathcal{E}(n)t}\|n\rangle\!\rangle\langle\!\langle n\|\psi(0)\rangle
\label {inisol} \\
&=\sum_{n,z\in{V}}e^{-i\mathcal{E}(n)t}\|n\rangle\!\rangle\hat{\phi}_n(z)\varphi(z),
\label {inisol2}
\end{align}
and the solution of the initial value problem of the discrete time BD process is given by
\begin{align} 
 |\psi^d(\ell)\rangle=U^\ell |\psi^d(0)\rangle
&=\sum_{n\in{V}}e^{-i\mathcal{E}^d(n)\ell}\|n\rangle\!\rangle\langle\!\langle n\|\psi^d(0)\rangle
\label {inisold} \\
&=\sum_{n,z\in{V}}e^{-i\mathcal{E}^d(n)\ell}\|n\rangle\!\rangle\hat{\phi}_n(z)\varphi^d(z).
\label {inisol2d}
\end{align}
The time evolution is causal and the non-causal effects of branching appear at the measurements.
For the initial state $|\psi^{(d)}(0)\rangle=|y\rangle$ ($\varphi^{(d)}(z)=\delta_{z\,y}$) 
at $t=0$,  ($\ell=0$), the probability amplitude of  arriving at 
the state $|x\rangle$ at time $t$ ($\ell$) is
\begin{align} 
\Psi(x,y;t)&\eqdef\langle x|e^{-i\mathcal{H}t}|y\rangle=
\sum_{n\in{V}}e^{-i\mathcal{E}(n)t}\langle x\|n\rangle\!\rangle\langle\!\langle n\|y\rangle
=\sum_{n\in{V}}e^{-i\mathcal{E}(n)t}\hat{\phi}_n(x)\hat{\phi}_n(y),
\label{psiform}\\
\Psi^d(x,y;\ell)&\eqdef\ \, \langle x|U^\ell|y\rangle\ =
\sum_{n\in{V}}e^{-i\mathcal{E}^d(n)\ell}\langle x\|n\rangle\!\rangle\langle\!\langle n\|y\rangle
=\sum_{n\in{V}}e^{-i\cE^d(n)\ell}\hat{\phi}_n(x)\hat{\phi}_n(y).
\label{psiformd}
\end{align}
The probability of measuring the state $|x\rangle$ at time $t$ ($\ell$)  is
\begin{align} 
|\Psi(x,y;t)|^2&=
\sum_{n\in{V}}\hat{\phi}_n(x)^2\hat{\phi}_n(y)^2
+2\!\!\sum_{n>m\in{V}}\!\!\cos[\bigl(\mathcal{E}(n)-\mathcal{E}(m)\bigr)t]
\hat{\phi}_n(x)\hat{\phi}_m(x)\hat{\phi}_n(y)\hat{\phi}_m(y),
\label{psi2}\\
|\Psi^d(x,y;\ell)|^2&=
\sum_{n\in{V}}\hat{\phi}_n(x)^2\hat{\phi}_n(y)^2
+2\!\!\sum_{n>m\in{V}}\!\!\cos[\bigl(\cE^d(n)-\cE^d(m)\bigr)\ell]
\hat{\phi}_n(x)\hat{\phi}_m(x)\hat{\phi}_n(y)\hat{\phi}_m(y).
\label{psi2d}
\end{align}
Various stochastic quantities can be calculated based on these  expressions.
Since Schr\"odinger  equation is time-reversal invariant, 
the above expressions do not converge  as $t\to\infty$ ($\ell\to\infty$).
In other words, there is no stationary (or terminal, limiting) distribution.
The long time average is 
\begin{align}
\lim_{T\to\infty}\frac1T\int_0^T|\Psi(x,y;t)|^2dt&=\sum_{n\in{V}}\hat{\phi}_n(x)^2\hat{\phi}_n(y)^2,
\label{taverage}\\
\lim_{T\to\infty}\frac1T\sum_{\ell=0}^{T}|\Psi^d(x,y;\ell)|^2&=\sum_{n\in{V}}\hat{\phi}_n(x)^2\hat{\phi}_n(y)^2,
\label{taverage2}
\end{align}
which is symmetric in $x$ and $y$.
\end{theo}
For the actual quantum systems in Nature, all excited states ($n\ge1)$ decay to lower energy states
as time goes on by the interactions with the environment.
The effect could be mimicked by adding a small imaginary part to the eigenvalues of the excited states;
\begin{equation*}
\mathcal{E}(n)\ \to  \mathcal{E}'(n)\eqdef
\left\{
\begin{array}{cc}
\mathcal{E}(n)-i\epsilon  &   n\ge1   \\
 0 &     n=0.
\end{array}
\right.\quad \epsilon >0.
\end{equation*}
Therefore the long time limit of the probability amplitude of the transition 
to the  state $|\chi\rangle=\sum_{u\in{V}}\chi(u)|u\rangle$ is
\begin{equation*}
\lim_{t\to\infty}\langle\chi|\psi(t)\rangle
=\lim_{t\to\infty}\sum_{n,u,z\in{V}}e^{-i\mathcal{E}'(n)t}\chi(u)^*\hat{\phi}_n(u)\hat{\phi}_n(z)\varphi(z)=\sum_{u,z\in{V}}\hat{\phi}_0(u)\hat{\phi}_0(z)\chi(u)^*\varphi(z).
\end{equation*}
For $\varphi(z)=\delta_{z\,y}$, $\chi(u)=\delta_{u\,x}$ this becomes
\begin{equation*}
\lim_{t\to\infty}\Psi(x,y;t)=\phi_0(x)\phi_0(y),\quad \lim_{t\to\infty}|\Psi(x,y;t)|^2=\phi_0(x)^2\phi_0(y)^2=\pi(x)\pi(y).
\end{equation*}
\begin{rema}
The marked similarity of the quantum probability amplitude $\Psi(x,y;t)$ \eqref{psiform} and the 
classical probability
$\mathcal{P}(x,y;t)$ \eqref{cpr1} was reported in (13) of \cite{childs2} and  (3.13), (2.21) of \cite{grunbaum}
and in different contexts in \cite{bessis, mantica}.
\end{rema}

%%%%%%%%%%%%%%%%%%%%%%%%%%%%%%%%%%%%%%%%%%%%%%%%%%%%%%%%%%%%%%%
%                                                             %
%  5. Explicitly solvable examples                              %
%                                                             %
%%%%%%%%%%%%%%%%%%%%%%%%%%%%%%%%%%%%%%%%%%%%%%%%%%%%%%%%%%%%%%%
\section{Explicitly solvable examples; Quantum cases}
\label{sec:exa}
The main purpose of this paper is to draw the experts' attention to {\bf 16 explicit examples} of 
{\em exactly solvable} continuous time BD processes so that detailed numerical comparison with the
classical versions could be made. Among them {\bf 11} provide  exactly solvable  examples of discrete time BD processes.
The corresponding exactly solved results of continuous time versions 
were published \cite{bdsol} more than ten years ago. The discrete time version was published \cite{dtbd} about a
year ago.
So only the quantum cases are reported here.

%%%%%%%%%%%%%%%%%%%%%%%%%%%%%%%%%%%%%%%%%%%%%%%%%%%%%
%                                                   %
% 5.1  Continuous time BD                       %
%                                                   %
%%%%%%%%%%%%%%%%%%%%%%%%%%%%%%%%%%%%%%%%%%%%%%%%%%%%%
\subsection{Continuous time BD }
\label{sec:cBDex}

The central idea is that for 16 specific choices of birth rate $B(x)$, death rate $D(x)$ and the eigenvalue
$\mathcal{E}(n)$, the eigenvalue equation \eqref{hteig1} of the tri-diagonal matrix 
$\widetilde{\mathcal H}$  \eqref{htdef}--\eqref{htdef3}
becomes identical with the difference equation 
\begin{align}
&B(x)\bigl(\check{P}_n(x)-\check{P}_n(x+1)\bigr)+D(x)\bigl(\check{P}_n(x)-\check{P}_n(x-1)\bigr)
=\mathcal{E}(n)\check{P}_n(x),
\quad x,n\in{V},
\label{difeq}
\end{align}
governing the hypergeometric orthogonal polynomials
of a discrete variable $\{\check{P}_n(x)\}$ belonging to the Askey scheme \cite{os12,askey,ismail,kls}.
These polynomials are (1) Krawtchouk, (2) Hahn, (3) dual Hahn, (4) Racah, (5) $q$-Krawtcouk, 
(6) quantum Krawtchouk, (7) affine $q$-Krawtchouk, (8) $q$-Hahn, (9) dual $q$-Hahn, 
(10) $q$-Racah for finite BD processes and (11) Charlier, (12) Meixner, (13) little $q$-Jacobi,
(14) little $q$-Laguerre, (15) Al-Salam-Carlitz II and (16) alternative $q$-Charlier for infinite BD processes.
For the details of these polynomials consult \cite{kls}, 
from which the naming is also adopted.
In a paper entitled ``Orthogonal polynomials from Hermitian matrices"  \cite{os12} by Odake and myself,
it is demonstrated that these polynomials are the eigenvectors of the Hamiltonians 
$\mathcal{H}$ \eqref{Hdef1}--\eqref{Hdef3} which later turn out to be corresponding to the
quantum BD processes of finite or infinite dimensions. 
Since all these polynomials also satisfy the normalisation condition \eqref{vnorm}
\begin{equation*}
\check{P}_n(0)=1,\quad n\in{V},
\end{equation*}
the eigenvectors of $\cH$ and $\widetilde{\cH}$ are
\begin{equation*}
\phi_n(x)=\phi_0(x)\check{P}_n(x),\quad n\in{V}.
\end{equation*}

%%%%%%%%%%%%%%%%%%%%%%%%%%%%%%%%%%%%%%%%%%%%%%%%%%%%%
%                                                   %
% 5.2  Discrete time BD                       %
%                                                   %
%%%%%%%%%%%%%%%%%%%%%%%%%%%%%%%%%%%%%%%%%%%%%%%%%%%%%
\subsection{Discrete time BD }
\label{sec:dBDex}

Since the discrete time BD Hamiltonian $\cH^d$ \eqref{Hddef}, \eqref{Hddef2} has the same form as the
continuous time BD Hamiltonian $\cH$ \eqref{Hdef1}--\eqref{Hdef3}, 
the exactly solvable discrete time BD is obtained by rescaling $B(x)$ and $D(x)$ of the exactly solvable
continuous time BD in terms of a  free parameter $t_S$  which sets the timescale of the discrete time BD,
\begin{align}
&b(x)\eqdef t_SB(x),\quad d(x)\eqdef t_SD(x),\quad 0<t_S\cdot\max_{x\in{V}}\bigl(B(x)+D(x)\bigr)<1,
\label{bdts}\\
&\Longrightarrow  \cH^d=t_S\cH,\quad \cE^d(n)=t_S\cE(n),\quad n\in{V}.
\label{cdErel}
\end{align}
The boundedness of $B(x)+D(x)$ excludes (11) Charlier, (12) Meixner, (13) little $q$-Jacobi,
(14) little $q$-Laguerre and and (16) alternative $q$-Charlier  among the infinite systems. 
For them $B(x)+D(x)$ is unbounded.
It should be stressed that rescaling $B(x)$ and $D(x)$ above does not change $\phi_0(x)$  \eqref{phi0def}
and $\pi(x)$ \eqref{pidef} since they are defined by the ratio $B(x)/D(x+1)$.

%%%%%%%%%%%%%%%%%%%%%%%%%%%%%%%%%%%%%%%%%%%%%%%%%%%%%
%                                                   %
% 5.3  Explicit Formulas                       %
%                                                   %
%%%%%%%%%%%%%%%%%%%%%%%%%%%%%%%%%%%%%%%%%%%%%%%%%%%%%
\subsection{Explicit formulas}
\label{sec:exform}

The main formulas in {\bf Theorem \ref{theo3}}, the amplitude $\Psi(x,y;t)$ \eqref{psiform}, 
$\Psi^d(x,y;t)$ \eqref{psiformd}, 
and the corresponding branching probability $|\Psi(x,y;t)|^2$ \eqref{psi2},  $|\Psi^d(x,y;t)|^2$ \eqref{psi2d}
are valid by redefining $\phi_n(x)$ \eqref{phindef} as
\begin{align}
\phi_n(x)&\eqdef \phi_0(x)\check{P}_n(x), \quad n\in{V},
\label{phindef3}\\
\Rightarrow \Psi(x,y;t)&={\phi}_0(x){\phi}_0(y)\sum_{n\in{V}}d_n^2e^{-i\mathcal{E}(n)t}
\check{P}_n(x)\check{P}_n(y)
\label{psiform2}\\
\Rightarrow \Psi^d(x,y;\ell)&={\phi}_0(x){\phi}_0(y)\sum_{n\in{V}}d_n^2e^{-i\mathcal{E}(n)t_S\ell}
\check{P}_n(x)\check{P}_n(y)
\label{psiform2d}\\
\Rightarrow |\Psi(x,y;t)|^2&={\phi}_0(x)^2{\phi}_0(y)^2\Bigl|\sum_{n\in{V}}d_n^2e^{-i\mathcal{E}(n)t}
\check{P}_n(x)\check{P}_n(y)\Bigr|^2,
\label{psi3}\\
\Rightarrow |\Psi^d(x,y;\ell)|^2&={\phi}_0(x)^2{\phi}_0(y)^2\Bigl|\sum_{n\in{V}}
d_n^2e^{-i\mathcal{E}(n)t_S\ell}
\check{P}_n(x)\check{P}_n(y)\Bigr|^2.
\label{psi3d}
\end{align}
In particular
\begin{align}
\Psi(x,0;t)&={\phi}_0(x)\sum_{n\in{V}}d_n^2e^{-i\mathcal{E}(n)t}
\check{P}_n(x),
\label{psiform3}\\
\Psi^d(x,0;\ell)&={\phi}_0(x)\sum_{n\in{V}}d_n^2e^{-i\mathcal{E}(n)t_S\ell}
\check{P}_n(x),
\label{psiform3d}
\end{align}
and 
\begin{align}
\Psi(N,0;t)&={\phi}_0(N)\sum_{n\in{V}}d_n^2e^{-i\mathcal{E}(n)t}
\check{P}_n(N),
\label{psiform4}\\
\Psi^d(N,0;\ell)&={\phi}_0(N)\sum_{n\in{V}}d_n^2e^{-i\mathcal{E}(n)t_S\ell}
\check{P}_n(N),
\label{psiform4d}
\end{align}
take simple forms as $\check{P}_n(0)=1$ and $\phi_0(0)=1$ and $\check{P}_n(N)$ is independent of $N$
 for (1) Krawtchouk, (2) Hahn, (5) $q$-Krawtchouk, (6) quantum Krawtchouk, (7) affine $q$-Krawtchouk 
 and (8) $q$-Hahn due to the
 $N$-dependence of the ($q$)-hypergeometric expressions.
These should be compared with the classical results
\begin{align}
\mathcal{P}(x,y;t)&={\phi}_0(x)^2
\sum_{n\in{V}}d_n^2e^{-\mathcal{E}(n)t}\check{P}_n(x)\check{P}_n(y),
\label{cpr2}\\
\mathcal{P}^d(x,y;\ell)&=\phi_0(x)^2
\sum_{n\in\mathcal{X}}d_n^2(1-t_S\cE(n))^\ell\check{P}_n(x)\check{P}_n(y),
\label{cpr2d}\\
\mathcal{P}(x,0;t)&={\phi}_0(x)^2
\sum_{n\in{V}}d_n^2e^{-\mathcal{E}(n)t}\check{P}_n(x),
\label{cpr3}\\
\mathcal{P}^d(x,0;\ell)&=\phi_0(x)^2
\sum_{n\in\mathcal{X}}d_n^2(1-t_S\cE(n))^\ell\check{P}_n(x),
\label{cpr3d}\\
\mathcal{P}(N,0;t)&={\phi}_0(N)^2
\sum_{n\in{V}}d_n^2e^{-\mathcal{E}(n)t}\check{P}_n(N),
\label{cpr4}\\
\mathcal{P}^d(x,0;\ell)&=\phi_0(x)^2
\sum_{n\in\mathcal{X}}d_n^2(1-t_S\cE(n))^\ell\check{P}_n(N),
\label{cpr4d}
\end{align}
which look similar to the quantum expressions.
In order to evaluate  the amplitude $\Psi(x,y;t)$ $\Psi^d(x,y;t)$ \eqref{psiform2}--\eqref{psiform4}
and the corresponding branching probability $|\Psi(x,y;t)|^2$ \eqref{psi3} and 
$|\Psi^d(x,y;t)|^2$ \eqref{psi3d}  explicitly and exactly for these 
16 different  BD processes,
the  data of $B(x)$, $D(x)$, $\mathcal{E}(n)$ and $\check{P}_n(x)$  \eqref{difeq}, $\phi_0(x)^2$ \eqref{pidef}, 
  and $d_n$ \eqref{orth}  are necessary.  They  are given in \S4 of the paper
``Exactly solvable birth and death processes," \cite{bdsol}.

It should be stressed  that  among these 16 polynomials
some  have all integer eigenvalues $\mathcal{E}(n)=n$, (1) Krawtchouk, 
(3) dual Hahn, (11) Charlier and (12) Meixner and some others have all integer eigenvalues for integer
parameters only $\mathcal{E}=n(n+d)$ type (2) Hahn and  (4) Racah.
For these cases, the {\em quantum theory}, in particular, $\Psi(x,y;t)$ \eqref{psiform2} or 
$\Psi^d(x,y;\ell)$ \eqref{psiform2d}
is {\em $2\pi$ periodic} like the harmonic oscillator with or without centrifugal force
(Laguerre and Hermite polynomial) for the all integer eigenvalues cases and the P\"oschl-Teller potential
(Jacobi polynomial) for the integer parameter cases.
Sometimes these periodic systems are said to exhibit perfect return.
The rest are $q$-polynomials and their spectra are more interesting.
Three of them, (7) affine $q$-Krawtchouk,  (8) $q$-Hahn and (14) little $q$-Laguerre, have the 
eigenvalues $\mathcal{E}(n)=q^{-n}-1$. 
Their quantum systems  have all integer eigenvalues and therefor become {\em $2\pi$ periodic} 
when  $q$ is chosen  
 $q=1/k$, $k\in{\mathbb N}\backslash\{1\}$. 
 These look extremely rapid propagation. 
 This is because the birth and death rates $B(x)$ and $D(x)$  are chosen very big, for example,
 $B(0)\sim N, q^{-N}$, in order to obtain simple forms of the eigenvalues $\mathcal{E}(n)\sim n, n^2, q^{-n}$ etc,
 see  \eqref{KBD}, \eqref{HBD}, \eqref{qHBD}, \eqref{qKBD} and \eqref{CBD} shown below.
 Comparison with the ordinary one-dimensional quantum systems would be helpful.
 A particle confined in a length $L$ box has eigenvalues $\mathcal{E}(n)\propto ({n}/{L})^2$.
 A pendulum of length $L$ has the frequency $\omega\propto 1/\sqrt{L}$, 
 meaning $\mathcal{E}(n)\propto n/\sqrt{L}$.
 This means that it would be natural to divide $B(x)$ and $D(x)$ by a factor of the order of $\sqrt{N}$
 for finite systems with the linear spectrum $\mathcal{E}(n)\sim n$ and divide by a factor of order $N^2$ 
 for finite systems with quadratic spectrum $\mathcal{E}(n)\sim n^2$.

 The rest has the spectra $\mathcal{E}(n)=1-q^n$,
 or $\mathcal{E}(n)=(q^{-n}-1)(1-pq^n)$.
If wanted, finite systems can be made periodic with   the period $q^{-N}2\pi$ by 
multiplying $B(x)$ and $D(x)$  a factor $q^{-N}$, $q=1/k$, $k\in\mathbb{N}\backslash\{1\}$.
For them the naturalness would require a proper adjusting factor. 

 For the discrete time quantum BD processes, the eigenvalues are bounded 
 $0\leq \cE^d(n)\le2$ \eqref{eigbound2}.
 If the exactly solvable continuous time BD is periodic, the corresponding discrete time BD is also
 periodic with the period equal to the continuous time period divided by $t_S$.
%%%%%%%%%%%%%%%%%%%%%%%%%%%%%%%%%%%%%%%%%%%%%%%%%%%%%
%                                                   %
% 5.1     Data of $B(x)$, $D(x)$ etc                   %
%                                                   %
%%%%%%%%%%%%%%%%%%%%%%%%%%%%%%%%%%%%%%%%%%%%%%%%%%%%%
\subsection{Data of $B(x)$, $D(x)$ etc }
\label{sec:data}

Here I present the data of a few polynomials to convey the vivid image of the problem setting and to avoid 
massive duplication.
As seen below the data have many free parameters in contradistinction to most quantum walks
which have virtually no free parameter.
The ranges of the parameters are restricted so that the birth rate 
$B(x)$ and the death rate $D(x)$ are positive.
No numerical evaluation of the time developments of the amplitudes $\Psi(x,y;t)$ \eqref{psiform2} 
and the corresponding branching probability $|\Psi(x,y;t)|^2$ \eqref{psi3} will be provided in this paper 
since  the situation is not ripe for a quick numerical result.
The data and some formulas of the probability amplitudes and branching probabilities of  four 
finite BD processes,  Krawtchouk,  Hahn, 
 $q$-Hahn, quantum $q$-Krawtchouk  and  one infinite BD process, Charlier are provided below. 
It is interesting to note that the amplitude $\Psi(x,y;t)$ \eqref{psiform}  from one end to the other
$\Psi(N,0;t)$ for some finite cases become very simple.
Throughout this section, the $N$ dependence in the notation of the finite polynomials and the
normalisation constant $d_n$ is suppressed
for simplicity of presentation.

%%%%%%%%%%%%%%%%%%%%%%%%%%%%%%%%%%%%%%%%%
\subsubsection{ Krawtchouk }
\label{Kr}
The case of linear (in $x$) birth and death rates is  a very well-known example
(the Ehrenfest model) \cite{KarMcG}
of an exactly solvable birth and death processes \cite{feller}:
\begin{align}
  B(x)&=p(N-x),\quad
  D(x)=(1-p)x,\quad 0<p<1,\quad  \mathcal{E}(n)=n,
  \label{KBD}\\
  \phi_0(x)^2&=
  \frac{N!}{x!\,(N-x)!}\Bigl(\frac{p}{1-p}\Bigr)^x,\quad
  d_n^2
  =\frac{N!}{n!\,(N-n)!}\Bigl(\frac{p}{1-p}\Bigr)^n\times(1-p)^N,\\[4pt]
    \check{P}_n(x)&=P_n(x)
  ={}_2F_1\Bigl(
  \genfrac{}{}{0pt}{}{-n,\,-x}{-N}\Bigm|p^{-1}\Bigr),\quad  
  \check{P}_n(N)={}_1F_0\Bigl(
  \genfrac{}{}{0pt}{}{-n}{-}\Bigm|p^{-1}\Bigr)=(1-p^{-1})^n.
\end{align}
The stationary probability 
$\pi(x)=\phi_0(x)^2d_0^2$ \eqref{pidef} of the classical system  is the binomial distribution.
The transition from $0$ to $N$ has simple expressions,
\begin{align} 
\Psi(N,0;t)=\bigl(p(1-p)\bigr)^{N/2}(1-e^{-it})^N \ 
\Rightarrow \bigl|\Psi(N,0;t)\bigr|^2=\bigl(p(1-p)\bigr)^N2^{2N}(\sin\tfrac{t}2)^{2N}.
\end{align}
The corresponding classical transition probability reads
\begin{equation}
\mathcal{P}(N,0;t)=p^N(1-e^{-t})^N.
\end{equation}
It is interesting to note that for $p=1/2$, $\Psi(N,0;(2n+1)\pi)=1$ independent of  the size $N$.
This is a very fast propagation due to the symmetry of the system. This is due to the property 
$(-1)^n\check{P}_n(x)=\check{P}_n(N-x)$
of Krawtchouk polynomial for $p=1/2$, as $\phi_0(N)=1$ and
\begin{equation*}
\Psi(N,0;(2n+1)\pi)=\sum_{n\in{V}}d_n^2=\sum_{n\in{V}}d_n^2\phi_n(0)^2=1.
\end{equation*}
Since all the eigenvalues are integers, the system is $2\pi$ periodic. 
That is, for $t=2m\pi$, $m\in{\mathbb Z}$, the time dependence of  \eqref{psiform2} drops and
\begin{align}
 \Psi(x,y;2m\pi)&={\phi}_0(x){\phi}_0(y)\sum_{n\in{V}}d_n^2
\check{P}_n(x)\check{P}_n(y)=\sum_{n\in{V}}\hat{\phi}_n(x)\hat{\phi}_n(y)=\delta_{x\,y}.
\label{perio1}
\end{align}
This applies to all the systems having all integer eigenvalues.
%%%%%%%%%%%%%%%%%%%%%%%%%%%%%%%%%%%%%%%%
\subsubsection{Hahn}
\label{Hn}
The birth and death rates are quadratic in $x$ and the eigenvalues are quadratic in $n$, too.
\begin{align}
B(x)&=(x+a)(N-x),\quad
  D(x)= x(b+N-x), \quad a,b>0,
  \label{HBD}\\
\mathcal{E}(n)&= n(n+a+b-1),\quad
\phi_0(x)^2
 =\frac{N!}{x!\,(N-x)!}\,\frac{(a)_x\,(b)_{N-x}}{(b)_N},\\
  d_n^2
  &=\frac{N!}{n!\,(N-n)!}\,
  \frac{(a)_n\,(2n+a+b-1)(a+b)_N}{(b)_n\,(n+a+b-1)_{N+1}}
  \times\frac{(b)_N}{(a+b)_N},\\[4pt]
  \check{P}_n(x)&=P_n(x)
  ={}_3F_2\Bigl(
  \genfrac{}{}{0pt}{}{-n,\,n+a+b-1,\,-x}
  {a,\,-N}\Bigm|1\Bigr),\\
 &\check{P}_n(N)={}_2F_1\Bigl(
  \genfrac{}{}{0pt}{}{-n,\,n+a+b-1}
  {a}\Bigm|1\Bigr)=\frac{(-1)^n(b)_n}{(a)_n}.
 \end{align}
 When $a+b$ is a positive integer the system is periodic and \eqref{perio1} holds. 
 For example, if  $a+b=2$, the period is $\pi$.
The transition amplitude from $0$ to $N$ reads
\begin{align} 
\Psi(N,0;t)=\sqrt{(a)_N(b)_N}\sum_{n\in{V}}\frac{N!}{n!(N-n)!}(-1)^n\frac{2n+a+b-1}{(n+a+b-1)_{N+1}}
e^{-in(n+a+b-1)t}.
\end{align}
The classical probability has a similar expression,
\begin{align} 
\mathcal{P}(N,0;t)=(a)_N\sum_{n\in{V}}\frac{N!}{n!(N-n)!}(-1)^n\frac{2n+a+b-1}{(n+a+b-1)_{N+1}}
e^{-n(n+a+b-1)t}.
\end{align}
%%%%%%%%%%%%%%%%%%%%%%%%%%%%%%%%%
\subsubsection{$q$-Hahn}
\label{qHn}

The birth and death rates are  quadratic  in $q^x$ and the eigenvalues are also quadratic in $q^n$,
\begin{align}
B(x)&=(1-aq^x)(q^{x-N}-1),\quad
  D(x)= aq^{-1}(1-q^x)(q^{x-N}-b),\quad 0<a,b<1,
  \label{qHBD}\\
 \mathcal{E}(n)
  &=(q^{-n}-1)(1-abq^{n-1}),\qquad
  \eta(x)=q^{-x}-1,\\
    \phi_0(x)^2
  &=\frac{(q\,;q)_N}{(q\,;q)_x\,(q\,;q)_{N-x}}\,
  \frac{(a;q)_x\,(b\,;q)_{N-x}}{(b\,;q)_N\,a^x}\,,\\[4pt]
  d_n^2
  &=\frac{(q\,;q)_N}{(q\,;q)_n\,(q\,;q)_{N-n}}\,
  \frac{(a,abq^{-1};q)_n}{(abq^N,b\,;q)_n\,a^n}\,
  \frac{1-abq^{2n-1}}{1-abq^{-1}}
  \times\frac{(b\,;q)_N\,a^N}{(ab\,;q)_N},\\[4pt]
   \check{P}_n(x)&=P_n(\eta(x))
  ={}_3\phi_2\Bigl(
  \genfrac{}{}{0pt}{}{q^{-n},\,abq^{n-1},\,q^{-x}}
  {a,\,q^{-N}}\Bigm|q\,;q\Bigr),\\
 & \check{P}_n(N)={}_2\phi_1\Bigl(
  \genfrac{}{}{0pt}{}{q^{-n},\,abq^{n-1}}
  {a}\Bigm|q\,;q\Bigr)=(-a)^nq^{n(n-1)/2}\frac{(b\,;q)_n}{(a\,;q)_n}.
\end{align}
The transition amplitude from $0$ to $N$ reads
\begin{align} 
\Psi(N,0;t)&=\frac{\sqrt{(a\,;q)_N(b\,;q)_Na^N}}{(ab\,;q)_N}
\sum_{n\in{V}}\frac{(q\,;q)_N}{(q\,;q)_n\,(q\,;q)_{N-n}}\,(-1)^n
  \frac{(abq^{-1};q)_n}{(abq^N\,;q)_n}\,
  \frac{1-abq^{2n-1}}{1-abq^{-1}}\n
  & \hspace{5cm}\times q^{n(n-1)/2}
e^{-i(q^{-n}-1)(1-abq^{n-1})t}.
\end{align}
The classical probability has a similar expression with $it\to t$,
\begin{align} 
\mathcal{P}(N,0;t)&=\frac{(a\,;q)_N}{(ab\,;q)_N}
\sum_{n\in{V}}\frac{(q\,;q)_N}{(q\,;q)_n\,(q\,;q)_{N-n}}\,(-1)^n
  \frac{(abq^{-1};q)_n}{(abq^N\,;q)_n}\,
  \frac{1-abq^{2n-1}}{1-abq^{-1}}\n
  & \hspace{5cm}\times q^{n(n-1)/2}
e^{-(q^{-n}-1)(1-abq^{n-1})t}.
\end{align}
%%%%%%%%%%%%%%%%%%%%%%%%%%%%%%%%%%%%
\subsubsection{quantum $q$-Krawtchouk}
This has a universally bounded spectrum $\mathcal{E}(n)=1-q^n<1$.
The birth and death rates are
quadratic polynomials in $q^x$:
\begin{align}
  B(x)&=p^{-1}q^x(q^{x-N}-1),\qquad
  D(x)=(1-q^x)(1-p^{-1}q^{x-N-1}),\quad p>q^{-N},
  \label{qKBD}\\
%
%  \mathcal{E}(n)&=1-q^n,\qquad
%  \eta(x)=q^{-x}-1,\\
%
  \phi_0(x)^2
  &=\frac{(q\,;q)_N}{(q\,;q)_x(q\,;q)_{N-x}}\,
  \frac{p^{-x}q^{x(x-1-N)}}{(p^{-1}q^{-N}\,;q)_x}\,,\quad   \eta(x)=q^{-x}-1,\\[4pt]
  d_n^2
  &=\frac{(q\,;q)_N}{(q\,;q)_n(q\,;q)_{N-n}}\,
  \frac{p^{-n}q^{-Nn}}{(p^{-1}q^{-n}\,;q)_n}\,
  \times(p^{-1}q^{-N}\,;q)_N,\\[4pt]
  \check{P}_n(x)&= P_n(\eta(x))
  ={}_2\phi_1\Bigl(
  \genfrac{}{}{0pt}{}{q^{-n},\,q^{-x}}{q^{-N}}\Bigm|q\,;pq^{n+1}\Bigr),\quad
  \check{P}_n(N)={}_1\phi_0\Bigl(
  \genfrac{}{}{0pt}{}{q^{-n}}{-}\Bigm|q\,;pq^{n+1}\Bigr)=(pq\,;q)_n.
\end{align}
The transition amplitude from $0$ to $N$ reads
\begin{align}
\Psi(N,0;t)&=\sqrt{\frac{(pq)^{-N}}{(p^{-1}q^{-N}\,;q)_N}}
\sum_{n\in{V}}\frac{(q\,;q)_N}{(q\,;q)_n(q\,;q)_{N-n}}\,
  \frac{p^{-n}q^{-Nn}}{(p^{-1}q^{-n}\,;q)_n}
(p^{-1}q^{-N}\,;q)_N\,e^{-i\mathcal{E}(n)t}(pq\,;q)_n\n
  &=\sqrt{(pq)^{-N}(p^{-1}q^{-N}\,;q)_N}\sum_{n\in{V}}\frac{(q\,;q)_N}{(q\,;q)_n(q\,;q)_{N-n}}\,
  (-1)^nq^{-Nn}q^{n(n+1)/2}e^{-i\mathcal{E}(n)t}.
\end{align}
The classical probability has a similar expression with $it\to t$,
\begin{align}
\mathcal{P}(N,0;t)&=(pq)^{-N}\sum_{n\in{V}}\frac{(q\,;q)_N}{(q\,;q)_n(q\,;q)_{N-n}}\,
  (-1)^nq^{-Nn}q^{n(n+1)/2}e^{-\mathcal{E}(n)t}.
\end{align}
%%%%%%%%%%%%%%%%%%%%%%%%%%%%%%%%%%
\subsubsection{Charlier}
\label{Ch}
Another best known example of exactly solvable birth and death process.
The data are
\begin{align}
  B(x)&=a,\qquad
  D(x)=x,\qquad a>0, \qquad 
  \mathcal{E}(n)=n,
  \label{CBD}\\[4pt]
  \phi_0(x)^2&=\frac{a^x}{x!}\,,\quad
  d_n^2
  =\frac{a^{n}}{n!}\times e^{-a}, \quad 
  \check{P}_n(x)= P_n(x)
  ={}_2F_0\Bigl(
  \genfrac{}{}{0pt}{}{-n,\,-x}{-}\Bigm|-a^{-1}\Bigr).
  \label{charlP}
\end{align}
The quantum system is $2\pi$ periodic.
The stationary probability $\pi(x)=\phi_0(x)^2d_0^2$ \eqref{pidef} of the classical system  is the Poisson distribution.
The probability amplitude for going from $0$ to $x$ is
\begin{equation}
\Psi(x,0;t)=\sqrt{\frac{a^x}{x!}}\sum_{n\in{V}}\frac{a^n}{n!}e^{-a-int}{}_2F_0\Bigl(
  \genfrac{}{}{0pt}{}{-n,\,-x}{-}\Bigm|-a^{-1}\Bigr).
\end{equation}
The classical probability has a similar expression,
\begin{equation}
\mathcal{P}(x,0;t)=\frac{a^x}{x!}\sum_{n\in{V}}\frac{a^n}{n!}e^{-a-nt}{}_2F_0\Bigl(
  \genfrac{}{}{0pt}{}{-n,\,-x}{-}\Bigm|-a^{-1}\Bigr).
\end{equation}

%%%%%%%%%%%%%%%%%%%%%%%%%%%%%%%%%%%%%%%%%%%%%%%%%%%%%
%                                                   %
% 6. Comments                           %
%                                                   %
%%%%%%%%%%%%%%%%%%%%%%%%%%%%%%%%%%%%%%%%%%%%%%%%%%%%%
\section{Comments}
\label{sec:comm}
A few comments are in order.
\begin{itemize}
\item  I follow the standard convention of quantum mechanics, 
$\langle \rm{final\ state}|e^{-i\mathcal{H}t}|{\rm initial\ state}\rangle$.
The corresponding probability $\mathcal{P}(x,y;t)$ also means the probability of starting at the initial state $y$ at time 0
and arriving at the  final state $x$  after time $t$. In probability community, the opposite convention is dominant.
\item The Karlin-McGregor representation \cite{KarMcG} is a well-known solution method of BD processes.
A brief comparison/contrast with my approach would be helpful.
The orthogonal polynomials play important roles in both approaches.
In  \cite{KarMcG}, the population is denoted by $n,m\in{V}$. The birth rate $\lambda_n$ and
death rate $\mu_n$ define orthogonal polynomials $\{Q_n(x)\}$ through three term recurrence relation
\begin{equation*}
-xQ_n(x)=\mu_nQ_{n-1}(x)-(\lambda_n+\mu_n)Q_n(x)+\lambda_nQ_{n+1}(x),
\end{equation*}
which is dual to the difference equation \eqref{difeq}.
Here $x$ is the spectral parameter. The positive orthogonality measure $\psi$ on $0\le x<\infty$, 
\begin{equation*}
\int_0^\infty Q_i(x)Q_j(x)d\psi(x)=0,\quad i\neq j,\tag{\cite{KarMcG}.2.4}
\end{equation*}
is determined through a certain moment problem. The probability from $j$ to $i$  after time $t$ is
\begin{equation*}
\mathcal{P}(i,j;t)=\pi_i\int_0^\infty e^{-xt}Q_i(x)Q_j(x)d\psi(x),\tag{\cite{KarMcG}.1.7}
\end{equation*}
in which $\pi_i$ is the same as $\pi(i)$ defined in \eqref{pidef}. 
In the approach no  (symmetric) matrix 
appears explicitly  whose spectrum is  related with $x$ and the measure $\psi$.
Such matrix  would be  a candidate for the quantum Hamiltonian.
I presume this is one of the reasons why there are relatively few papers discussing quantum BD processes.
\item Gr\"unbaum, Vinet and Zhedanov \cite{grunbaum},  formulated continuous time quantum BD processes following 
Karlin-McGregor approach. They adopted a Hamiltonian $\mathcal{H}'$ which is related to $\mathcal{H}$ 
\eqref{LBDHrel}--\eqref{Hdef3} by a similarity transformation
\begin{equation*}
\mathcal{H}'_{x\,y}=(-1)^x\mathcal{H}_{x\,y}(-1)^y,
\end{equation*}
which has all positive matrix elements. They also showed an interesting connection between a 
continuous time quantum BD process
$\cH'$ and a continuous time spin chain with single excitation state,
whose Hamiltonian ${\cH'}^{spin}$ reads in the notation of this paper
\begin{align*}
{\cH'}^{spin}&\eqdef \frac12\sum_{x\in{V}}\sqrt{B(x)D(x+1)}
\bigl((\sigma^1)_x(\sigma^1)_{x+1}+(\sigma^2)_x(\sigma^2)_{x+1}\bigr)\\
&\ +\frac12\sum_{x\in{V}}\bigl(B(x)+D(x)\bigr)\bigl((\sigma^3)_x+1).  \tag{\cite{grunbaum}.3.2}
\end{align*}
Here $(\sigma^j)_x$ $j=1,2,3$ is the Pauli sigma matrix at vertex $x$.
The solution of the spin problem is obtained by the solution of the eigenvalue problem of the matrix $\cH'$.
This means all exactly solvable BD processes reported in the present  paper and \cite{bdsol} provide as many exactly 
solvable spin chains both continuous and discrete time. And many of them are periodic.
\item Some  exact analytical results of continuous time quantum walks on graphs  are worth mentioning.
They are  on a cycle graph \cite{solenov}, on a Cayley tree \cite{mulken}, on a ring with $N$ vertices and each vertex is connected with $m$ 
nearest neighbours on the right and left \cite{xu1} and on a star graph \cite{xu2}.
\item Time averaged distributions of quantum walks were reported in many papers.
It should be stressed that these results are initial state specific and not universal, cf \eqref{taverage}.
\item The present solution method does not apply to $q$-Meixner and $q$-Charlier, although they are listed in
\S4 of \cite{bdsol}. This is because these polynomials fail to satisfy the completeness relation \eqref{phincomp}
as proven in \cite{atakishi}. These two polynomials and others which have Jackson integral measures
provide another type of exactly solvable classical BD processes \cite{os34}. 
It is a good challenge to formulate their quantum versions. 
\item
The same solution techniques apply to quantum BD processes related to 
various new orthogonal polynomials \cite{os23}. 
They are deformations of  the classical orthogonal polynomials, {\em e.g.} the Racah and $q$-Racah, etc
 obtained by multiple applications of the discrete analogue of the Darboux transformations or 
the Krein-Adler transformations. These new polynomials offer virtually infinite examples of 
exactly solvable quantum BD processes.
\end{itemize}

%%%%%%%%%%%%%%%%%%%%%%%%%%%%%%%%%%%%%%%%%%%%%%%%%%%%%%%%%%%%%%%
%                                                             %
%  Acknowledgments                                            %
%                                                             %
%%%%%%%%%%%%%%%%%%%%%%%%%%%%%%%%%%%%%%%%%%%%%%%%%%%%%%%%%%%%%%%
\section*{Acknowledgements}
RS thanks Mourad Ismail for inducing him to explore the mysterious maze of birth and death processes.
%%%%%%%%%%%%%%%%%%%%%%%%%%%%%%%%%%%%%%%%%%%%%%%%%%%%%%%%%%%%%%%
%                                                             %
%  Appendix A.                                                %
%                                                             %
%%%%%%%%%%%%%%%%%%%%%%%%%%%%%%%%%%%%%%%%%%%%%%%%%%%%%%%%%%%%%%%
\section*{\hspace{7cm}Appendix \\
Basic  definitions related to the  ($q$)-hypergeometric functions}
\label{appendA}
\setcounter{equation}{0}
\renewcommand{\theequation}{A.\arabic{equation}}

%Some basic  definitions related to
%the ($q$-)hypergeometric functions are listed here \cite{kls}.

\noindent
%%%%%%%%%%%%%%%%%%%
$\circ$ shifted factorial $(a)_n$ :
\begin{equation}
  (a)_n\eqdef\prod_{k=1}^n(a+k-1)=a(a+1)\cdots(a+n-1)
  =\frac{\Gamma(a+n)}{\Gamma(a)}.
  \label{defPoch}
\end{equation}
%%%%%%%%%%%%%%%%%%%
$\circ$ $q$-shifted factorial $(a\,;q)_n$ :
\begin{equation}
  (a\,;q)_n\eqdef\prod_{k=1}^n(1-aq^{k-1})=(1-a)(1-aq)\cdots(1-aq^{n-1}).
  \label{defqPoch}
\end{equation}
%%%%%%%%%%%%%%%%%%%
$\circ$ hypergeometric function ${}_rF_s$ :
\begin{equation}
  {}_rF_s\Bigl(\genfrac{}{}{0pt}{}{a_1,\,\cdots,a_r}{b_1,\,\cdots,b_s}
  \Bigm|z\Bigr)
  \eqdef\sum_{n=0}^{\infty}
  \frac{(a_1,\,\cdots,a_r)_n}{(b_1,\,\cdots,b_s)_n}\frac{z^n}{n!}\,,
  \label{defhypergeom}
\end{equation}
where $(a_1,\,\cdots,a_r)_n\eqdef\prod_{j=1}^r(a_j)_n
=(a_1)_n\cdots(a_r)_n$.\\
%%%%%%%%%%%%%%%%%%%
$\circ$ $q$-hypergeometric function (the basic hypergeometric function)
${}_r\phi_s$ :
\begin{equation}
  {}_r\phi_s\Bigl(
  \genfrac{}{}{0pt}{}{a_1,\,\cdots,a_r}{b_1,\,\cdots,b_s}
  \Bigm|q\,;z\Bigr)
  \eqdef\sum_{n=0}^{\infty}
  \frac{(a_1,\,\cdots,a_r\,;q)_n}{(b_1,\,\cdots,b_s\,;q)_n}
  (-1)^{(1+s-r)n}q^{(1+s-r)n(n-1)/2}\frac{z^n}{(q\,;q)_n}\,,
  \label{defqhypergeom}
\end{equation}
where $(a_1,\,\cdots,a_r\,;q)_n\eqdef\prod_{j=1}^r(a_j\,;q)_n
=(a_1\,;q)_n\cdots(a_r\,;q)_n$.

%%%%%%%%%%%%%%%%%%%%%%%%%%%%%%%%%%%%%%%%%%%%%%%%%%%%%%%%%%%%%%%
%                                                             %
%  References                                                 %
%                                                             %
%%%%%%%%%%%%%%%%%%%%%%%%%%%%%%%%%%%%%%%%%%%%%%%%%%%%%%%%%%%%%%%


\begin{thebibliography}{99}
%author alphabetical order

\bibitem{aharonov}
D.\,Aharonov, A.\,Ambainis, J.\,Kempe and U.\,Vazirani,
 ``Quantum walks on graphs," In Proceedings of the 33th ACM Symposium on The Theory of Computation (STOC’01) ACM, 
(2001) 50--59, 
{\tt arXiv:quant-ph/0012090}.


\bibitem{ambainis}
A.\,Ambainis, J.\,Kempe and A.\, Rivosh, ``Coins make quantum walks faster," Proc. 16th
ACM-SIAM SODA, (2005) 1099--1108,  {\tt arXiv:quant-ph/0402107}.

\bibitem{ambainis2}
A.\,Ambainis,  E.\,Bach,  A.\,Nayak, A.\,Vishwanath and  J.\,Watrous,
``One-Dimensional Quantum Walks,"
In Proceedings of the 33rd ACM Symposium on The Theory of Computation
(STOC’01) ACM, (2001) 60--69.

\bibitem{aldous}
D.\,Aldous and J.\,A.\,Fill, ``Reversible Markov Chains and Random Walks on Graphs,"
 in preparation, http://www.stat.berkeley.edu/~aldous/RWG/book.html.

\bibitem{askey}
G.\,E.\,Andrews, R.\,Askey and R.\,Roy,
{\it Special Functions},
%vol. 71 of Encyclopedia of mathematics and its applications,
Encyclopedia of mathematics and its applications,
Cambridge Univ. Press, Cambridge, (1999).

\bibitem{atakishi}
M.\,N.\,Atakishiyev, N.\,M.\,Atakishiyev and A.\,U.\,Klimyk,
``Big $q$-Laguerre and $q$-Meixner polynomials and representation of the
algebra $U_q(su_{1,1}$),''
J. Phys. {\bf A36} (2003) 10335-10347,
{\tt arXiv:math/0306201[math.QA]}.



% open system
\bibitem{attal}
S.\,Attal, F.\,Petruccione, C.\,Sabot and I.\,Sinayskiy,
``Open Quantum Random Walks,"
Journal of Statistical Physics {\bf 147}  (2012) 832--852,
{\tt arXiv:1402.3253[quant-ph]}.

\bibitem{bessis}
D.\,Bessis and G.\,Mantica, ``Orthogonal polynomials associated to almost periodic Schrijdinger
operators. A trend towards random orthogonal polynomials," J. Comp. Appl. Math. {\bf 48} ( 1993)
17-32.

 \bibitem{chase}
 B.\,A.\,Chase and A.\,J.\,Landhal,  ``Universal quantum walks and adiabatic algorithms by 1d hamiltonians," 
 {\tt arXiv:0802.1207[quant-ph]}.


\bibitem{childs}
 A.\,Childs, E.\,Farhi and S.\,Gutmann,  ``An example of the difference between quantum and classical random walks,"
  Quantum Information Processing, {\bf 1} (2002) 35--43,
  {\tt arXiv:\hspace{0pt}quant-ph/0103020}.
  
  \bibitem{childs2}
  A.\,Childs, 
  ``On the relationship between continuous- and discrete-time quantum walk,"
  Comm. Math. Phys. {\bf 294} (2010) 581--603,
  {\tt arXiv:0810.0312[quant-ph]}.


\bibitem{diaconis20}
P.\,Diaconis and C.\,Zhong,
``Hahn polynomials and Burnside process,''
Ramanujan J. https://doi.org/10.1007/s11139-021-00482-z,
{\tt arXiv:2012.\hspace{0pt}13829[math.PR]}.



\bibitem{farhi}
E.\,Farhi and S.\,Gutmann, ``Quantum computation and decision trees,"
Phys. Rev. {\bf A58}  (1998) 915--928, 
{\tt arXiv:quant-ph/9706062}.


\bibitem{feller}
W.\,Feller,
``The birth and death processes as diffusion processes,''
J. Math. Pures Appl. (9) {\bf 38} (1959) 301--345.
W.\,Feller, {\it An Introduction to Probability Theory and its Applications,
I}, (2nd ed.), Wiley, New York, (1966).

\bibitem{albert}
F.\,A.\,Gr\"unbaum and M.\,Rahman,
``A System of Multivariable Krawtchouk Polynomials and a Probabilistic
Application,''
SIGMA {\bf 7} (2011) 119,
{\tt arXiv:1106.1835[math.PR]}.

\bibitem{grunbaum}
 A.\,F.\,Gr\"unbaum, L.\,Vinet and  A.\,Zhedanov,
``Birth and death processes and quantum spin chains,"
J. Math. Phys. {\bf 54}  (2013) 062101,
{\tt arXiv:1205.4689[quant-ph]}.  


\bibitem{ho}
C.-L.\,Ho, Y.\,Ide, N.\,Konno, E.\,Segawa and K.\,Takumi, 
``A spectral analysis of discrete-time quantum walks with related to birth and death chains,” 
J. Stat. Phys. {\bf 171}  (2018) 207--219, {\tt  arXiv:1706.01005 [quant-ph]}.


\bibitem{hoa-rah83}
M.\,R.\,Hoare and M.\,Rahman,
``Cumulative Bernoulli trials and Krawtchouk processes,''
Stochastic Processes and their applications {\bf 16} (1983) 113--139.



\bibitem{ismail}
M.\,E.\,H.\,Ismail,
{\it Classical and Quantum Orthogonal Polynomials in One Variable\/},
Encyclopedia of mathematics and its applications,
Cambridge Univ. Press, Cambridge, (2005).



\bibitem{KarMcG}
S.\,Karlin and J.\,L.\,McGregor, ``The differential equations
of birth-and-death processes," Trans. Amer. Math. Soc. {\bf 85} (1957)
489--546.

\bibitem{KarMcG2}
S.\,Karlin and J.\,L.\,McGregor, 
``The classification of birth and death processes,"
Trans. Am. Math. Soc. {\bf 86} (1957) 366--400.

\bibitem{kempe}
J.\,Kempe,  ``Quantum random walks - an introductory overview,"
Contemporary Physics {\bf 44} (2003) 307--327,
{\tt arXiv:quant-ph/0303081}. 



\bibitem{kls}
R.\,Koekoek, P.\,A.\,Lesky and R.\,F.\,Swarttouw,
{\it Hypergeometric orthogonal polynomials and their $q$-analogues,\/}
Springer Monographs in Mathematics,
Springer-Verlag Berlin-Heidelberg, (2010).

\bibitem{konno}
N.\,Konno,
``Quantum Random Walks in One Dimension,"
 Quantum Information Processing  {\bf 1} (2002)  345--354. https:/\!/doi.org/10.1023/A:1023413713008,
 {\tt arXiv:quant-ph\hspace{0pt}/0206053}.


 \bibitem{mantica}
 G.\,Mantica, ``Quantum intermittency in almost-periodic lattice systems derived from their 
 spectral properties,"  Physica {\bf D103} (1997) 576-589.


 \bibitem{meyer}
 D.\,A.\,Meyer,  ``From quantum cellular automata to quantum lattice gases," 
 J. Stat. Phys.,
{\bf 85} (1996) 551--574, {\tt arXiv:quant-ph/9604003}.


\bibitem{mulken}
O.\,M\"ulken and A.\,Blumen, ``Slow transport by continuous time quantum walks,"
 Phys. Rev. {\bf E71} (2005) 016101,
{\tt arXiv:quant-ph/0410243};
``Spacetime structures of continuous time quantum walks,"
Phys. Rev. {\bf E71} (2005) 036128, 
{\tt arXiv:quant-ph/052004}.

\bibitem{mulken2}
O.\,M\"ulken and A.\,Blumen,  ``Continuous-time quantum walks: Models for coherent 
transport on complex networks," Physics Reports {\bf 502} (2011)  37--87, 
{\tt arXiv:1101.\hspace{0pt}2572[quant-ph]}.

\bibitem{nayak}
A.\,Nayak and A.\,Vishwanath,  ``Quantum walk on the line," {\tt arXiv:quant-ph/0010117}.


\bibitem{os12}
S.\,Odake and R.\,Sasaki,
``Orthogonal Polynomials from Hermitian Matrices,''
J. Math. Phys. {\bf 49} (2008) 053503 (43 pp),
{\tt arXiv:0712.4106[math.CA]}.

\bibitem{os23}
S.\,Odake and R.\,Sasaki,
``Exceptional ($X_\ell$) ($q$)-Racah polynomials,"
Prog. Theor. Phys. {\bf 125} (2011) 851-870,
 {\tt arXiv:1102.0812[math-ph]};
 ``Multi-indexed ($q$-)Racah polynomials,"
J. Phys. {\bf A45} (2012) 385201 (21 pp),
{\tt arXiv:1203.5868[math-ph]};
``Multi-indexed Meixner and little $q$-Jacobi (Laguerre) Polynomials,"
J. Phys. {\bf A50} (2017) 165204 (23pp),
{\tt arXiv:1610.09854[math-ph]}. 

\bibitem{os34}
S.\,Odake and R.\,Sasaki,
``Orthogonal Polynomials from Hermitian Matrices \II,''
J. Math. Phys. {\bf 59} (2018) 013504 (42pp),
{\tt arXiv:1604.00714[math.CA]}.

\bibitem{os39}
S.\,Odake and R.\,Sasaki,
``Markov chains generated by convolutions of orthogonality measures,"
J. Phys. A: Math. Theor. {\bf 55} (2022) 275201 (42pp), {\tt arXiv:2106.04082\hspace{0pt}[math.PR]}.


\bibitem{patel}
A.\,Patel, K.\,S.\,Raghunathan and P.\,Rungta,  ``Quantum random walks do not need a coin
toss," Phys. Rev. {\bf A71}   (2005) 032347, {\tt arXiv:quant-ph/0405128}.



\bibitem{bdsol}
R.\,Sasaki,
``Exactly Solvable Birth and Death Processes,''
J. Math. Phys. {\bf 50} (2009) 103509 (18 pp),
{\tt arXiv:0903.3097[math-ph]}.


\bibitem{dtbd}
R.\,Sasaki,
``Exactly solvable discrete time Birth and Death processes,''
J. Math. Phys. {\bf 63}  (2022) 063305; https:/\!/doi.org/10.1063/5.0079228,
{\tt arXiv:2106.\hspace{0pt}03284[math.PR]}.

\bibitem{qcMarkov}
R.\,Sasaki,
``Quantum vs classical Markov chains; Exactly solvable examples,''
In preparation. {\tt arXiv:2212.xxxx[quant-ph]}.


 \bibitem{solenov}
 D.\,Solenov and L.\,Fedichkin, ``Continuous-time Quantum Walks on a Cycle Graph,"
 Phys. Rev. {\bf A75} (2007) 030301R, {\tt arXiv:quant-ph/0506096}.

\bibitem{szegedy}
M.\,Szegedy,  ``Quantum speed-up of Markov chain based algorithms,"
Proc. of the 45th Annual IEEE Symposium
on Foundations of Computer Science (FOCS’04), 32--41 (2004).



\bibitem{venegas}
S.\,E.\,Venegas-Andraca, ``Quantum walks: a comprehensive review," Quantum Inf. Process. {\bf 11}
(2012) 1015--1106, {\tt arXiv:1204.4780[quant-ph]}.


 \bibitem{watrous}
 J.\,Watrous, ``Quantum simulations of classical random walks and undirected graph connectivity,"
  J. Comput. System Sci. {\bf 62} (2001) 376--391, {\tt arXiv:cs/9812012}.


%classical
\bibitem{woess}
W.\,Woess, ``Random walks on infinite graphs and groups,"
Cambridge tracts in mathematics  {\bf 138}, Cambridge University Press, (2000).

\bibitem{xu1}
 X.-P.\,Xu,  
``Continuous-time quantum walks on one-dimensional regular networks,"
Phys. Rev. {\bf E77} (2008) 061127 , {\tt arXiv:0801.4180[quant-ph]}.

\bibitem{xu2}
 X.-P.\,Xu,  ``Exact analytical results for quantum walks on star graphs," J. Phys. A: Math.
Theor., {\bf 42} (2009)115205,   {\tt arXiv:0903.1149[quant-ph]}.


%
\end{thebibliography}
\end{document}